\documentclass{aa} 

\usepackage{graphicx}	
\usepackage{changepage}

\usepackage{bm}
\usepackage{txfonts}
\usepackage{amsmath}	
\usepackage{amssymb}	
\usepackage{amsfonts}	
\usepackage{tabularx}
\usepackage{hyperref}

\newcommand{\beq}{\begin{equation}}
\newcommand{\eeq}{\end{equation}}
\newcommand{\beqa}{\begin{eqnarray}}
\newcommand{\eeqa}{\end{eqnarray}}
\newcommand{\eqn}[1]{Eq.\,(\ref{#1})}
\newcommand{\fig}[1]{Fig.\,\ref{#1}}
\newcommand{\tab}[1]{Tab.\,\ref{#1}}
\newcommand{\sect}[1]{Sect.\,\ref{#1}}
\newcommand{\appx}[1]{Appendix\,\ref{#1}}
\newcommand{\dust}[1]{{\rm #1_d}}
\newcommand{\dusti}[1]{{\rm #1}_{{\rm d}, i}}
\newcommand{\dustj}[1]{{\rm #1}_{{\rm d}, j}}

\newcommand{\ith}{$i$th}
\newcommand{\jth}{$j$th}
\newcommand{\dd}{{\rm d}}
\newcommand{\npsym}{%
  \mathrel{\ooalign{\raisebox{.6ex}{\tiny min}\cr\raisebox{-.6ex}{\tiny max}}}
}

\def\lsim{\mathrel{\rlap{\lower 3pt \hbox{$\sim$}} \raise 2.0pt \hbox{$<$}}}
\def\gsim{\mathrel{\rlap{\lower 3pt \hbox{$\sim$}} \raise 2.0pt \hbox{$>$}}}
\def\msun{\rm {M_\odot}}

\begin{document}

\title{A novel framework to study the impact of binding energy distributions on the chemistry of dust grains}

\subtitle{}

\author{
    T. Grassi\inst{1,2,}\thanks{E-mail: tgrassi@usm.lmu.de}
    \and
    S. Bovino\inst{3} 
    \and
    P. Caselli\inst{4}
    \and
    G. Bovolenta\inst{5}
    \and
    S. Vogt-Geisse\inst{5}
    \and
    B. Ercolano\inst{1,2}
}

\institute{
    Universit\"ats-Sternwarte M\"unchen, Scheinerstr. 1, D-81679 M\"unchen, Germany
    \and
    Excellence Cluster Origin and Structure of the Universe, Boltzmannstr.2, D-85748 Garching bei M\"unchen, Germany
    \and
    Departamento de Astronom\'ia, Facultad Ciencias F\'isicas y Matem\'aticas, Universidad de Concepci\'on,\\
    Av. Esteban Iturra s/n Barrio Universitario, Casilla 160, Concepci\'on, Chile
    \and
    Max-Planck-Institut f\"{u}r extraterrestrische Physik, Giessenbachstrasse 1, D-85748 Garching, Germany
    \and
    Departamento de F\'isico-Qu\'imica, Facultad de Ciencias Qu\'imicas, Universidad de Concepci\'on, Concepci\'on, Chile
}

\date{Accepted XXX. Received YYY; in original form ZZZ}

\abstract{
The evaporation of molecules from dust grains is crucial to understand some key aspects of the star- and the planet-formation processes.
During the warm-up phase the presence of young protostellar objects induces molecules to evaporate from the dust surface into the gas phase, enhancing its chemical complexity. Similarly, in circumstellar disks, the position of the so-called snow-lines is determined by evaporation, with important consequences for the formation of planets. The amount of molecules that are desorbed depends on the interaction between the species and the grain surface, which is controlled by the binding energy. Recent theoretical and experimental works point towards a distribution of values for this parameter instead of the single value often employed in astrochemical models.We present here a new ``multi-binding energy'' framework, to assess the effects that a distribution of binding energies has on the amount of species bound to the grains. We find that the efficiency of the surface chemistry is significantly influenced by this process with crucial consequences on the theoretical estimates of the desorbed species.}


\maketitle


\section{Introduction}
Since \citet{Hasegawa1992,Hasegawa1993} the desorption process, i.e.~the evaporation of molecules from the surface of dust grains, has been modelled with a classic Polanyi-Wigner approach, where instantaneous desorption at a given binding energy\footnote{Binding energy $E_{\rm b}$ and binding temperature are related by the Boltzmann constant $k_{\rm B}$. The terms ``binding energy'' and ``binding temperature'' are used in this paper interchangeably.} $E_{\rm b}=k_{\rm B}T_{\rm b}$ is controlled by the rate $k_\mathrm{e}\propto\exp{(-T_{\rm b}/T_\mathrm{d})}$, with $T_{\rm d}$ being the temperature of the dust. This is somehow limiting our understanding of the evaporation process and, more important, might affect the interpretation of observational data through astrochemical models.

The desorption process \textit{per se} has been studied both theoretically \citep[e.g.][]{Fayolle2016,Penteado2017,Wakelam2017,das2018,Shimonishi2018,Romero} as well as experimentally \citep[e.g.][]{Collings2004,ISAC2010,Dulieu2013,Surf2014,Potapov2017,Tehule2019}. For example, temperature programmed desorption (TPD) experiments under different conditions provided binding energies as a function of coverage and substrate material \citep[e.g.][]{He21601,Noble2012,He2016}. However, these experiments also show some limitations, in particular due to sensitivity problems related to the measurements  and identification of the volatiles (through  mass spectrometry), and the difficulty to study radicals species (see the discussion in \citealt{Schlemmer2001} and the recent attempts to present a non-destructive detection method for the desorbed species e.g.~\citealt{Tehule2019} and \citealt{Yocum2019}).

On the other hand, theoretical studies have never been conducted systematically; most of them have employed idealized set-ups, considering for instance the interaction of the molecule of interest with a single water molecule \citep[see e.g.][]{Wakelam2017}, even though the energetics strongly depends on the geometrical configuration of the molecules embedded in a cluster (or in a typical solid structure  as the Amorphous Solid Water, ASW).

Some attempts at improvement have been pursued by \citet{das2018} who performed calculations of the binding energy of 100~molecules interacting with small clusters of up to six water molecules, and \citet{Shimonishi2018} that provided molecular dynamics simulations to properly describe a cluster of 20~water molecules and its interactions with carbon, nitrogen, and oxygen atoms. Recent works have shown improvements, \citep[see e.g.][]{Romero} but a systematic study, which mixes accurate molecular dynamics simulations and robust quantum chemistry methods is still missing. In a recent effort, \citet{Bovolenta2020} have built a robust pipeline to compute the binding energy of hydrogen fluoride (HF) on ASW showing a Gaussian-like distribution of the binding energy of the interacting sites, pointing out that the binding energy does not, in fact, have a single value. This will be extended in the future to study more molecules on realistic substrates by performing at the same time accurate molecular dynamics simulations and applying \textit{ab initio} methods to evaluate the energetics of such systems.

The few available experiments show that molecules interact with the surface of grains in different ways depending on the available type of sites \citep[e.g.][]{Watanabe2010,He2016}. Some sites are indeed more suitable for strong interactions and are usually the first to be populated, while ``peak'' sites (as opposed to ``valley'' sites) produce weaker interactions. If we consider the inverse process, i.e.~evaporation, the capability of a molecule to remain bound to the surface will be determined by the binding energies; if this indeed is not represented by a single value but rather by a Gaussian distribution \citep{He21601,Noble2012,Bovolenta2020}, the amount of molecules residing on the surface of grains could be larger than the one that assumes no distribution, since there are sites where molecules are bounded for longer times due to their greater binding energy.

This becomes relevant when modelling for example the chemistry of star-forming regions and protoplanetary discs \citep[see][and references therein]{Cuppen2017}, where evaporation is a crucial process, in particular for the formation of interstellar complex organic molecules (iCOMs) and for the position of the so-called snow-lines, i.e.~the region of a protoplanetary disk where volatiles evaporate from dust grains (see e.g.~\citealt{Stevenson1988,Zhang2015}). Models and theoretical studies currently fail to reproduce the observed chemical complexity reflected in the richness of rotational spectra seen in young stellar objects \citep[for an extensive review see][]{COMreview2020}. While gas-phase routes are now extensively studied \citep{Skouteris2018,Skouteris2019}, most of the astrochemical models still focus on the formation of these molecules on the surface of grains via thermal hopping, tunnelling, and other interactions \citep[][to cite some of the most recent]{Bonfand2019,Ruaud2019,Jin2020}. The chemistry of these molecules depends on the amount of available reactants on the surface during the warm-up phase: if their residence time is relatively short, their abundances will rapidly decrease and quenching the reactivity on the surface of the grain.

Over the last three decades, the development of more realistic and sophisticated models for dust surface chemistry has been mainly based on a ``multi-layer'' approach \citep[e.g.][]{Taquet2014,Vasyunin2017} rather than ``multi-binding''. While multi-layering is paramount to understand the reactivity on the surface of grains and the adsorption process, a multi-binding approach is crucial to determine the final amount of tracers which are released back into gas-phase, and then to provide a more realistic comparison with observations. The effect of varying the binding energy as a parameter following the available experiments has been explored for example by \citet{Taquet2014} and \citet{Penteado2017}, but without modelling a distribution. To the best of our knowledge, the only attempt to include multiple binding energies in the same chemical model has been pursued by e.g.~\citet{He2016}, but limited to the reactions relative to their specific experiments.

In this Paper, we propose a new framework to take into account the multi-binding nature of the gas-grain interactions by modifying the classical single-binding approach, as discussed in \sect{sect:methods}. In \sect{sect:models} we report some results and show the impact of the multi-binding approach on the surface chemistry by evolving the abundances of a chemical network. We finally present our conclusions in \sect{sect:conclusions}.

\section{Methods}\label{sect:methods}
\subsection{Single-binding energy framework}
In this work we consider three types of grain chemical reactions (see e.g.~\citealt{Cuppen2017}), namely freeze-out ($\rm X \to \dust{X}$), that is the sticking of a gas-phase species onto a dust grain, evaporation ($\rm \dust{X} \to X$), the inverse process, and formation/destruction reactions on the surface via the Langmuir-Hinshelwood diffusive mechanism ($\dust{X} + \dust{Y} \to products$).
The species involved in these reactions are controlled by the following differential equations
\beqa
    \dot n_{\rm X} &=& -R_{\rm f, X} + R_{\rm e, X} + \mathcal{C}_{\rm X}\\
    \dot n_{\rm \dust{X}} &=& R_{\rm f, X} - R_{\rm e, X} + \mathcal{H}_{\rm X}\,,
\eeqa
where $\mathcal{C}_{\rm X}$ contains all the formation and destruction reactions for X in the gas phase and $\mathcal{H}_{\rm X}$ all the formation and destruction reactions for X on the grain surface.

The freeze-out reaction rate for a grain is
\beq\label{eqn:freezeout}
    R_{\rm f, X} = \pi a^2 n_{\rm X} n_{\rm d} \mathrm{v}_{\rm X} S\,,
\eeq
where $\pi a^2$ is the grain geometrical cross-section with $a$ the grain size, $n_{\rm X}$ the volume density of the species in the gas phase, $n_{\rm d}$ the grain number density, $\mathrm{v}_{\rm X}$ the thermal speed of the species X
\beq
    \mathrm{v}_{\rm X} = \sqrt{\frac{8k_{\rm B} T}{\pi m_{\rm X}}}\,,
\eeq
where $k_{\rm B}$ is the Boltzmann constant, $T$ the temperature of the gas, and $m_{\rm X}$ the mass of X. The sticking coefficient $S$ represents the efficiency of the above process \citep{Hollenbach1979}
\beq
    S = \left[ 1 + 4\times10^{-2} \sqrt{T+T_{\rm d}} +2\times10^{-3} T + 8\times10^{-6} T^2 \right]^{-1}\,,
\eeq
with $T_{\rm d}$ the dust temperature (note that improved and recommended state-of-the-art sticking factors as e.g.~\citealt{He2016b}, do not affect the findings of our study).

\eqn{eqn:freezeout} can be easily generalized for a grain size distribution with $\varphi\propto a^p$, $p=-3.5$ \citep{Mathis1977}, defined in the range $a_{\rm min}$ to $a_{\rm max}$, with dust-to-gas mass ratio $\mathcal{D}$, and bulk density $\rho_0$, as
\beq\label{eqn:freezout_mrn}
 R_{\rm f, X} = n_{\rm X}\frac{\rho_{\rm g}\mathcal{D}}{4/3\rho_0}\frac{a^{p+3}}{a^{p+4}} \frac{p+4}{p+3}\mathrm{v}_{\rm X} S\,,
\eeq

Analogously, thermal desorption is controlled by the Polanyi-Wigner rate (e.g.~\citealt{Stahler1981,Grassi2017})
\beq\label{eqn:evaporation}
    R_{\rm e, X} = n_{\rm \dust{X}} \nu_0 \exp\left(-\frac{T_{\rm b, X}}{T_{\rm d}}\right)\,,
\eeq
where $\nu_0=10^{12}$\,s$^{-1}$ is the Debye frequency\footnote{In principle this number varies with the properties of the specific molecule, but the value employed here (and by other authors) does not affect our conclusions.} \citep{Tielens1987} and $E_{\rm b, X}=k_{\rm B}T_{\rm b,X}$ the binding energy of the species X on the grain site.

Surface reactions that belong to $\mathcal{H}_{\rm X}$, e.g.\, between $\dust{X}$ and $\dust{Y}$, are determined by the thermal hopping of the molecules on the surface (e.g.~\citealt{Hocuk2015,Cuppen2017})
\beq\label{eqn:reaction}
    R_{\rm r, X, Y} = \frac{n_{\rm \dust{X}}n_{\rm \dust{Y}}}{n_{\rm s}} \nu_0 P_{\rm b} \left[\exp\left(-g\frac{T_{\rm b, X}}{T_{\rm d}}\right) + \exp\left(-g\frac{T_{\rm b, Y}}{T_{\rm d}}\right)\right]\,,
\eeq
where $g=2/3$ and the number density of binding sites follows the same approach as \eqn{eqn:freezout_mrn}
\beq
    n_{\rm s} = 3\frac{\rho_{\rm g}\mathcal{D}}{\rho_0 \delta_{s}^2}\frac{a^{p+3}}{a^{p+4}} \frac{p+4}{p+3}\,,
\eeq
with a binding sites average distance $\delta_{\rm s} = 3$\,\AA, and tunnelling probability of crossing the rectangular barrier $E_{\rm a}$ of width $a_{\rm b}=1$\,\AA{} \citep{Hocuk2015}
\beq
    P_{\rm b} = \exp\left(-\frac{2a_{\rm b}}{\hbar}\sqrt{2\mu_{\rm X,Y} E_{\rm a}}\right)\,,
\eeq
where $\hbar$ is the reduced Planck constant and $\mu^{-1}_{\rm X,Y} = m_{\rm X}^{-1} + m_{\rm Y}^{-1}$ the reduced mass of the two species involved.

\subsection{Multi-binding energy framework}\label{sect:multi_binding}
The previous expressions hold until we assume that the binding sites, instead of having a unique binding energy $E_{\rm b, X}=k_{\rm B}T_{\rm b, X}$ per species, have different binding temperatures that follow a Gaussian distribution centred in $T_{\rm b, X}$ with variance $\sigma_{\rm X}^2$
\beq\label{eqn:gaussian}
    P(T_{\rm b}) = C \exp\left[-\frac{\left(T_{\rm b} - T_{\rm b, X}\right)^2}{2\sigma_{\rm X}^2}\right]\,,
\eeq
where $C$ is defined by the constraint
\beq
    C\int_{T_{\rm b, min}}^{T_{\rm b, max}} P(x)\, \dd x= 1\,,
\eeq
with $T_{\rm b, min}$ and $T_{\rm b, max}$ found by defining a lower limits ${\varepsilon=10^{-5}}$ of the Gaussian distribution that gives
\beq\label{eqn:tminmax}
    T_{\rm b, \npsym} = T_{\rm b, X} \mp \sigma_{\rm X}\sqrt{-2\ln\left(\varepsilon\right)}\,.
\eeq

The classic single-binding approach is the limiting case when \mbox{$\sigma_{\rm X}\to0$}. It is worth noting that this theoretical definition of the distribution can be replaced by more realistic distributions obtained by experiments and theoretical works \citep[see e.g.][]{He21601,Noble2012,Romero}, however in this paper we always assume a Gaussian distribution controlled by $T_{\rm b}$ and $\sigma_{\rm X}$, being compatible with some of the experimental findings so far, as well as easier to interpret within the assumptions/limitations of this work. The Gaussian becomes less accurate when for example the surface coverage is about to reach 1~monolayer, the binding energy approaches the value for multi-layers, and the corresponding cut-off in the energy value does not necessarily occur in the tail of $P$.

We discretise the binding energies for $\dust{X}$ and $\dust{Y}$ with $N_{\rm b}$ equally- and linearly-spaced bins (grains sites) in the range defined by \eqn{eqn:tminmax}, thus increasing the number of dust species by a factor $2\times N_{\rm b}$, and obtaining a new set of reactions
\beqa
    \rm X &\to& \dusti{X}\label{eqn:new_freezout}\\
    \rm Y &\to& \dusti{Y}\\
    \dusti{X} &\to& \rm X\\
    \dusti{Y} &\to& \rm Y\\
    \dusti{X} + \dustj{Y}  &\to& products\label{eqn:odexy}\,,
\eeqa
where $i$ represents the species on dust bound with the binding temperature in the \ith{} bin, i.e.~$T_{{\rm b}, i}$ (analogously for the \jth{} bin). The abundance $n_{\dust{X}}$ of the species $\dust{X}$ on the grain surface will be replaced by $N_{\rm b}$ abundances $n_{\dust{X}, i}$. Each freeze-out and evaporation reaction consists now of $N_{\rm b}$ reactions, for a total of $4\times N_{\rm b}+N_{\rm b}^{1+M}$ reactions, where the last term is due to \eqn{eqn:odexy}, with $M$ the number of products with multiple binding sites.

The new system of differential equations then reads
\beq
  \left\{
    \begin{aligned}
    \dot n_{\rm X} &=& - \sum_{i=1}^{N_{\rm b}} R_{{\rm f, X}, i} + \sum_{i=1}^{N_{\rm b}} R_{{\rm e, X},i} + \mathcal{C}_{\rm X}\\
    \dot n_{\rm Y} &=& - \sum_{j=1}^{N_{\rm b}} R_{{\rm f, Y}, j} + \sum_{j=1}^{N_{\rm b}} R_{{\rm e, Y},j} + \mathcal{C}_{\rm Y}\\
    \dot n_{\rm \dusti{X}} &=& R_{{\rm f, X}, i} - R_{{\rm e, X},i} - \sum_{j=1}^{N_{\rm b}} R_{{\rm X, Y}, i, j}\\
    \dot n_{\rm \dustj{Y}} &=& R_{{\rm f, Y}, j} - R_{{\rm e, Y},j} - \sum_{i=1}^{N_{\rm b}} R_{{\rm X, Y}, i, j}\,,\label{eqn:odey}
    \end{aligned}
  \right.
\eeq
where the rates are
\beqa
    R_{{\rm f, X}, i} &=& R_{\rm f, X} P(T_{{\rm b, X},i})\label{eqn:rate_f_x}\\
    R_{{\rm e, X}, i} &=& n_{\rm \dusti{X}} \nu_0 \exp\left(-\frac{T_{{\rm b, X},i}}{T_{\rm d}}\right)\label{eqn:rate_e_x}\\
    R_{{\rm X, Y}, i, j} &=& \frac{n_{\rm \dusti{X}} n_{\rm \dustj{Y}}}{n_{\rm s}}\nu_0 P_{\rm b}\nonumber\\
       &\times& \left[\exp\left(-g\frac{T_{{\rm b, X}, i}}{T_{\rm d}}\right) + \exp\left(-g\frac{T_{{\rm b, Y}, j}}{T_{\rm d}}\right)\right]\label{eqn:rate_xy}\,,
\eeqa
and the analogous to \eqn{eqn:rate_f_x} and \eqn{eqn:rate_e_x} for Y.

This restricted set of reactions already shows that a simple chemical network when $N_{\rm b}\gtrsim 10$ (see \sect{sect:parameters}) could be represented by a number of differential equations that is difficult to handle even by state-of-the-art differential equation integrators.
Reducing the computational cost of this approach is beyond the objectives of this Paper, however it could be possible to select some specific reactions that need to be ``expanded'' with a multi-binding approach, depending on what are the relevant chemical species that need to be tracked.

\eqn{eqn:rate_f_x} to \eqn{eqn:rate_xy} do not include interactions between bins of the same species (e.g.~$\dusti{X} \to \dustj{X}$). One of the limitations of a Gaussian with ``non-intercommunicating'' bins is that experiments show that the molecules tend to fill the sites with stronger bindings first and during the warming up molecules diffuse into different sites before desorption. This limitation can be overcome by including a diffusion term for molecules among different binding sites, with the drawback of increasing the number of rates by at least a factor $N_{\rm b}(N_{\rm b}-1)/2$ per molecule, assuming that the coefficients are available.

\section{Results}\label{sect:models}
In order to explore the effects of the multi-binding scenario, we have developed a dedicated and publicly-available\footnote{\url{https://bitbucket.org/tgrassi/multi\_bind}, \texttt{commit: 87c8a72}} \textsc{Python} framework that, given a chemical network in text form, writes the necessary code of the corresponding differential equations, rates, and Jacobian while running (i.e.~without the need of any pre-processor stage, as in e.g.~\textsc{krome} \citealt{Grassi2014}). The core of the code is \textsc{scipy}'s \textsc{solve\_ivp}, an implicit multi-step variable-order BDF solver \citep{Shampine1997}, that has a good balance between efficiency and ease of implementation. Our code also includes the pipeline for the analysis of the results.

We limit the set of reactions to the H-C-O chemical network\footnote{Chemical reactions are listed in \appx{sect:chemical_network}, rate coefficients can be found at \url{https://bitbucket.org/tgrassi/multi\_bind/src/master/networks/}.} from \citet{Glover2010} and \citet{Grassi2017} with the addition of the following surface reactions
\beqa
    \rm H_2O &\rightleftarrows& \dusti{H_2O}\\
    \rm CO &\rightleftarrows& \dusti{CO}\\
    \rm H &\rightleftarrows& \dusti{H}\\
    \dusti{H} + \dustj{CO} &\to& products\,,
\eeqa
where the subscripts $i$ and $j$ indicate that each one of the $N_{\rm b}$ binding energy bins includes that type of reaction.
The last reaction (with activation energy $E_{\rm a}/k_{\rm B}=2500\,{\rm K}$, \textsc{KIDA} database, \citealt{Wakelam2017}) is a key surface mechanism in prestellar cores (e.g.~\citealt{Vasyunin2017}) that leads to the formation of relevant molecules as H$_2$CO an CH$_3$OH by subsequent H-atom additions (e.g.~\citealt{Linnartz2015}), and here used as proxy to determine the efficiency of the process when changing $N_{\rm b}$.

\begin{figure}
 \includegraphics[width=\linewidth]{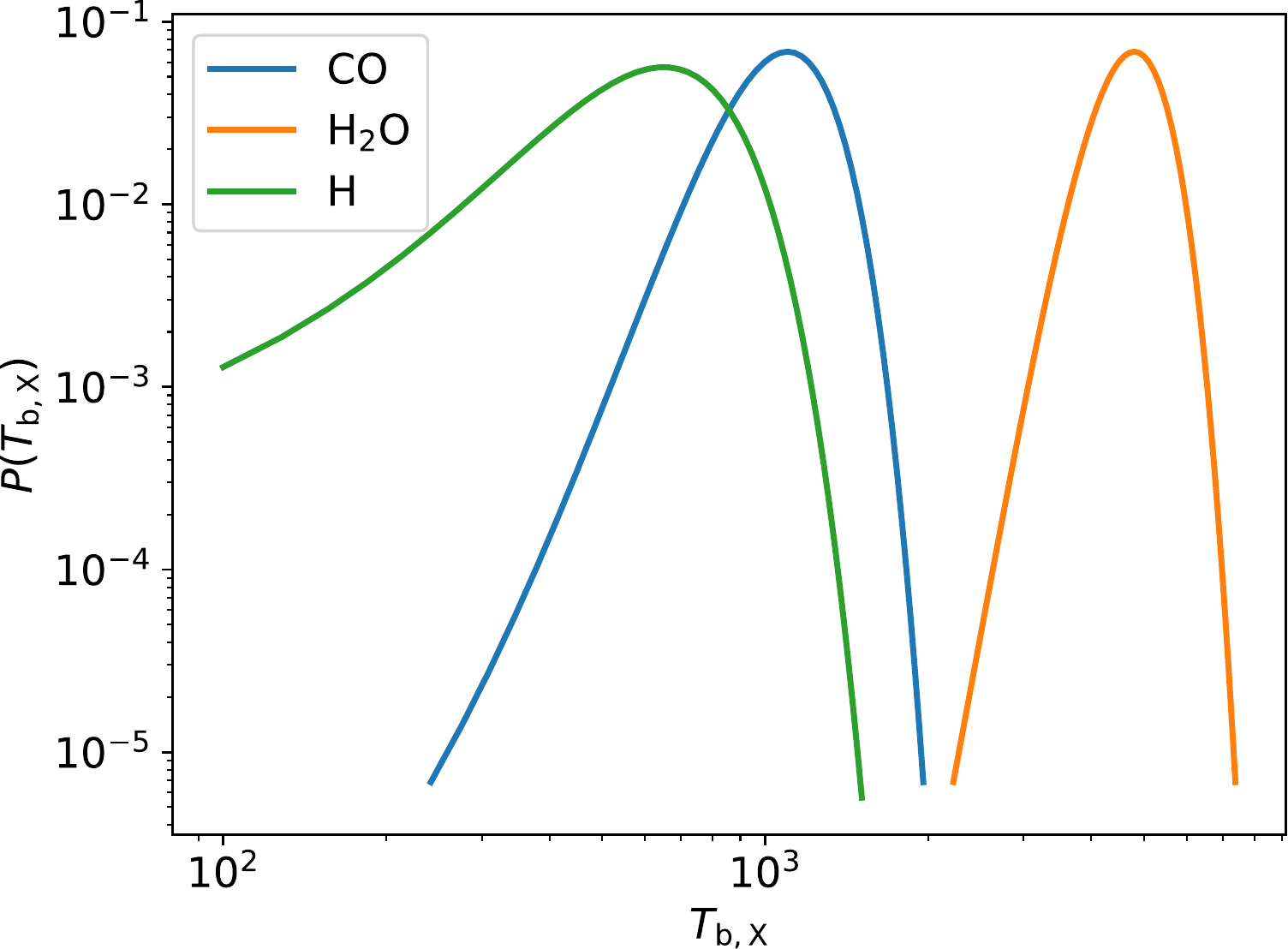}
 \caption{Distribution of the Gaussian binding temperatures $P(T_{\rm b, X})$ as defined in \eqn{eqn:gaussian} for CO (blue), water (orange), and H (green). Parameters $T_{\rm b, X}$ and $\sigma_{\rm X}$ are in \tab{tab:gauss}. Note the log-log scale.}\label{fig:p_distr}
\end{figure}

In this Paper we employ $N_{\rm b}=51$\,bins, and the $T_{\rm b, X}$ and $\sigma_{\rm X}$ values reported in \tab{tab:gauss} and plotted in \fig{fig:p_distr}, unless specified otherwise (see parameter sensitivity in \sect{sect:parameters}).  These values are compatible with the theoretical and experimental findings, but we do not refer to any specific experiment or theoretical calculation. However, being the values employed realistic, our conclusions are unaffected by the very specific choice (see also \sect{sect:parameters})

The dust grains have a size distribution $n_{\rm d}(a)\propto a^{p}=a^{-3.5}$ from $a_{\rm min}=5\times10^{-7}$\,cm to $a_{\rm max}=2.5\times10^{-5}$\,cm, dust-to-gas mass ratio $\mathcal{D}=10^{-2}$, and bulk density $\rho_0=3$\,g\,cm$^{-3}$, see \eqn{eqn:freezout_mrn}.

The rest of the chemical network is based on \citet{Glover2010} as employed by e.g.~\citet{Grassi2017}, and -- being far from completeness -- it is included here to test our framework without additional uncertain chemical processes that might complicate the process of analysis of the results. Despite this, our conclusions are independent from the chemical network employed.

\begin{table}
    \centering
	\begin{tabular}{lrr}
		\hline
        Species & $T_{\rm b, X}$/K & $\sigma_{\rm X}$/K\\
		\hline
        H & 650 & 200\\
        CO & 1100 & 200\\
        H$_2$O & 4800 & 600\\
		\hline
	\end{tabular}
	\caption{Mean $T_{\rm b, X}$ and standard deviation $\sigma_{\rm X}$ employed for the Gaussian functions in \eqn{eqn:gaussian}, that represent the distribution of the binding energies of the binding sites on the dust grains. See also \fig{fig:p_distr}.}\label{tab:gauss}
\end{table}

\subsection{Case study 1: The region surrounding a protostar with time-dependent luminosity evolution}\label{sect:protostar}
\begin{figure}
 \includegraphics[width=\linewidth]{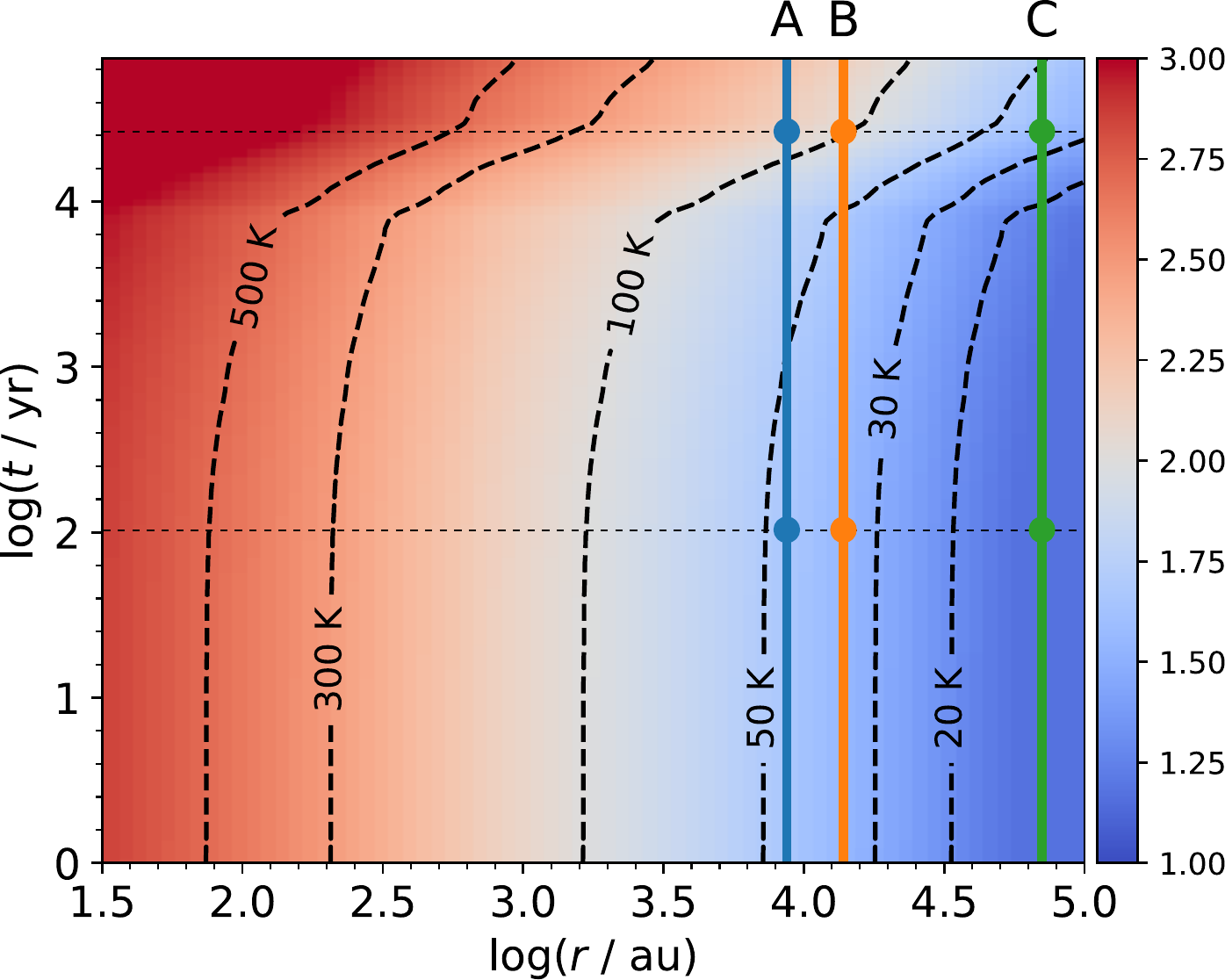}
 \caption{Temperature map as a function of $r$ and $t$ of the physical model calculated with the radiative transfer code \textsc{mocassin}, assuming $T=T_{\rm d}$, and where the colorbar reports $\log(T)$. The chemical network is evolved in time at each radius, varying the dust and the gas temperature according to this model, while density is a function of $r$ only. We selected three models at three specific radii (marked A, B, and C) to discuss the impact of the multiple binding energy approach, see \fig{fig:flux_diff}. At these radii we further discuss the distribution of the chemical abundances at specific $(r, t)$ combinations, indicated by the circular markers and by the horizontal dashed lines, see \fig{fig:flux_map}. The inner ($r\lesssim 3\times10^3$\,au) high-temperature region of the envelope is less relevant for the overall discussion, given the relatively short evaporation time-scale, and for this reason is ignored in our discussion.}\label{fig:tdust_map}
\end{figure}

In order to explore the effects of the variability of the dust temperature and of the density, we employ a physical model representing a clump of gas and dust around a high-mass protostar whose luminosity evolves in time following \citet{Stahler2005} and \citet{Hosokawa2009}.
The gas density profile \citep{Tafalla2002} is $n(r) = n_{\rm c}  r_{\rm c}^{2.5}\left(r_{\rm c}^{2.5}+r^{2.5}\right)^{-1}$,
where\footnote{We tested our model by changing $n_{\rm c}$ in the range $10^4$ to $10^7$\,cm$^{-3}$ and $r_{\rm c}$  in the range $10^4$ to $10^5$\,au, but their role in affecting our findings is negligible when compared to the role played by the variation in luminosity, the latter having the largest impact on the temperature profiles.} $n_{\rm c}=10^5$\,cm$^{-3}$, $r_{\rm c}=10^5$\,au, $r$ in au, and the dust mass density profile is $\rho_{\rm d}(r) = \mathcal{D}\, n(r)\mu\, m_{\rm p}$, where $\mu=2.34$ is the mean molecular weight. We compute the dust temperature profile by using the radiative transfer code \textsc{mocassin} \citep{Ercolano2003,Ercolano2005}, assuming that the protostar at the centre of the clump accretes mass with a rate of $\dot M=10^{-3}$\,$\msun$\,yr$^{-1}$ and we relate the mass of the protostar $M_*$ at a given time to its luminosity by employing the findings from \citet{Hosokawa2009}, their Fig.\,4. The temperature map found with this procedure is reported in \fig{fig:tdust_map}. Further information about the model are in Grassi~et~al., in preparation.

This physical model determines the density $n(r)$ and the temperature profile $T(r,t)=T_{\rm d}(r,t)$. At each radius we initialise the abundances of the species as in \citet{Rollig2007}, see \tab{tab:initial_abund}, and we evolve the system\footnote{The code to reproduce this model can be found at \url{https://bitbucket.org/tgrassi/multi\_bind/src/master/main.py}} assuming $n=10^4$\,cm$^{-3}$, $T=T_{\rm d}=10$\,K, for a time corresponding to the free-fall time at the given $r$. The cosmic-ray ionization rate is $\zeta=5\times10^{-17}$\,s$^{-1}$ and the visual extinction $A_{\rm v}=30$\,mag. The abundances obtained with this initial stage are then scaled by a factor $n(r)/10^4$\,cm$^{-3}$, and the chemical species are evolved with the time-dependent gas and dust temperature profiles obtained with the radiative transfer and shown in \fig{fig:tdust_map}.

\begin{figure}
 \includegraphics[width=\linewidth]{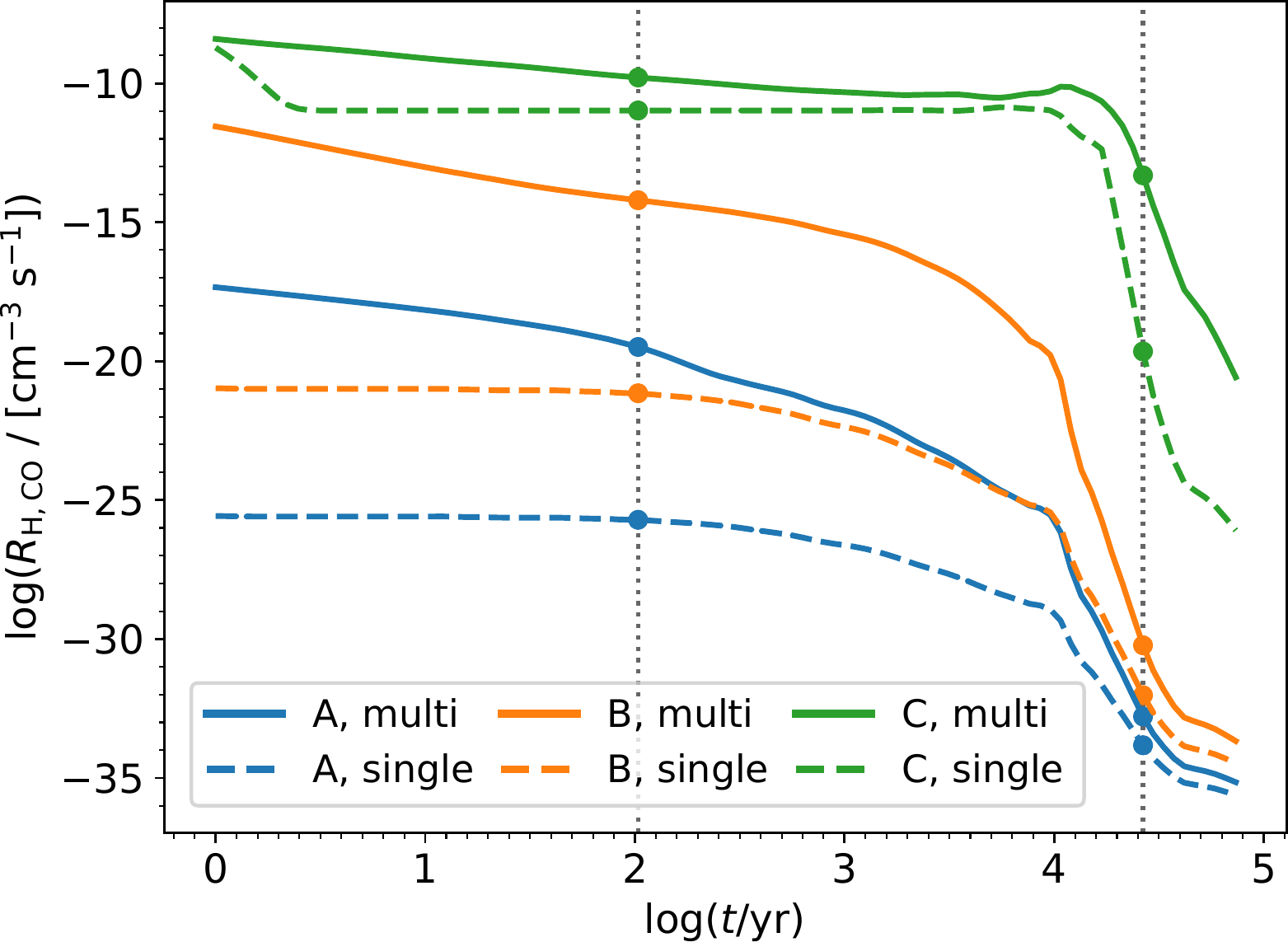}
 \caption{Time evolution of the rate $R_{\rm H, CO}$ obtained by summing the rates of the reactions that involve the individual bins with different binding energies, as defined in \eqn{eqn:flux_H_CO}. Note the decline of the rates with time, given by the increasing $T_{\rm d}$ that evaporates more and more reactants from the surface of the grains. The models are A (blue), B (orange), and C (green), as indicated in \fig{fig:tdust_map}, while solid and dashed lines indicate multiple- and single-bin approach, respectively. The vertical dotted lines and the circular markers are the same as in \fig{fig:tdust_map}, employed as a reference for \fig{fig:flux_map}.}\label{fig:flux_diff}
\end{figure}

In \fig{fig:flux_diff} we report the rate of $\dust{H} + \dust{CO} \to products$, defined by
\beq\label{eqn:flux_H_CO}
  R_{\rm H, CO}=\sum_{i=1}^{N_{\rm b}} \sum_{j=1}^{N_{\rm b}}R_{{\rm H, CO}, i,j}=\sum_{i=1}^{N_{\rm b}} \sum_{j=1}^{N_{\rm b}} k_{{\rm H, CO}, i, j}\,n_{\dust{H}, i}\,n_{\dust{CO}, j}\,,
\eeq
for the three different models indicated in \fig{fig:tdust_map} (namely A, B, and C), with the classical single-binding ($N_{\rm b}=1$) and with the new multi-binding ($N_{\rm b}=51$) approach. This rate is employed as a proxy to probe the efficiency of the process, and to determine the potential impact on the abundances of the different chemical species.

\fig{fig:flux_diff} shows a general decreasing of the flux with time, since in all the models the temperature increases simultaneously in time. Model A (``hot'') and C (``cold'') present respectively the smallest and the largest values of $R_{\rm H, CO}$ for both single- and multi-binding cases, with model B (``warm'') in between them.
The overall behaviour is determined by the abundances of the species ($n_{\dust{H}}$ and $n_{\dust{CO}}$) on the surface of the dust, that is proportional to the grain temperature and to their binding energy.

Similarly, this explains the difference between the single- and the multi-binding results for each model; given the distribution of binding sites, the latter includes also binding sites with higher binding energies that are capable of retaining the chemical species for longer times, and hence remaining available for the surface chemical reaction. The reaction flux in the latter case is orders of magnitude larger when compared to the single-binding energy  approach.

It is worth noticing that the binding energy distribution does not only have an effect on the abundances, but also on the rate coefficient $k_{\rm H, CO}$. In particular, the rate coefficient is controlled by the sum of the exponential hopping terms of the two reaction partners, see \eqn{eqn:reaction}, where higher $T_{\rm b}$ values reduce the mobility of the chemical species, quenching $k_{\rm H, CO}$. However, their sum is dominated by the term with lowest binding energy, hence $k_{\rm H, CO}$ is maximized when \emph{at least} one of the reaction partners belongs to a low binding energy site. On the contrary, when the dust temperature increases  the mobility of species on the surface increases, so that also at high-temperatures reactions will take place involving sites with higher binding energies. 

\begin{figure*}
 \includegraphics[width=\linewidth]{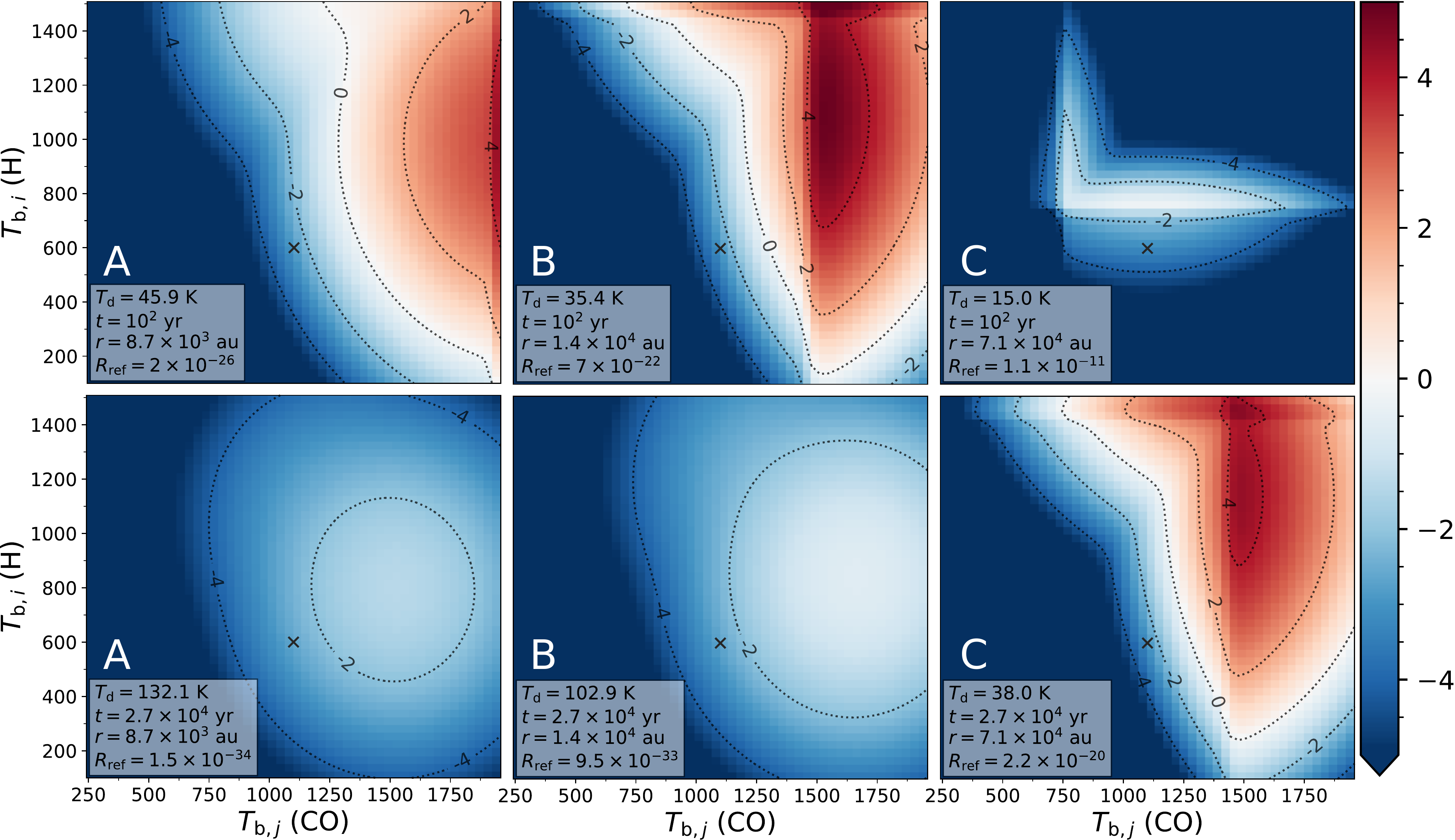}
 \caption{Logarithm of the ratio of $R_{{\rm H, CO},i,j}$ with the corresponding $R_{\rm ref}=R_{\rm H, CO}$ calculated with the single binding energy approach, for the $(r, t)$ combinations indicated with circular markers in \fig{fig:tdust_map} and \fig{fig:flux_diff}, namely cases A, B, and C (left to right panels), for $t=10^2$\,yr and $t=2.7\times10^4$\,yr (upper and lower panels). In the box we report the dust temperature $T_{\rm d}(r, t)$, $t$, $r$, and $R_{\rm ref}$ in units of cm$^{-3}$\,s$^{-1}$. The cross indicates the position of the single-case binding energy. For the sake of clarity the colour palette lower limit is set to $-5$. Compare with the sketch in \fig{fig:sketch_flux}.}\label{fig:flux_map}
\end{figure*}

\begin{figure}
 \includegraphics[width=\linewidth]{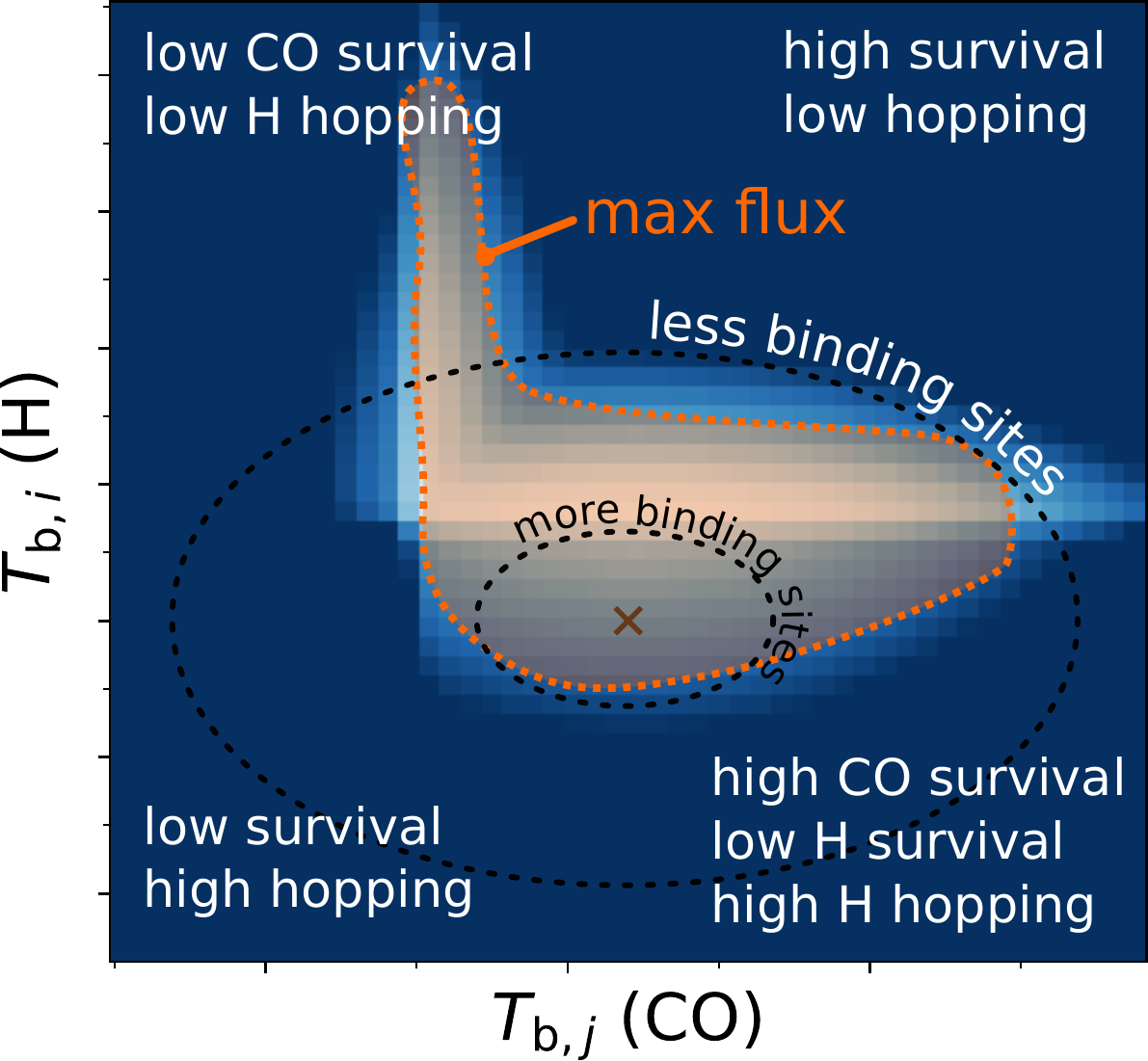}
 \caption{Schematic representation of the upper-right panel of \fig{fig:flux_map}, but also applicable to the other panels. Labels indicate what are the specific conditions that characterize the flux in those regions. Dashed circular lines represent the reduction of available binding sites due to the Gaussian shape of their distribution. The orange-shaded area is where the flux is maximized, due to the interplay of the above conditions. }\label{fig:sketch_flux}
\end{figure}

This interactions can be explained by showing the maps in \fig{fig:flux_map} (also sketched in \fig{fig:sketch_flux}), where we report the logarithm of the ratio between $R_{{\rm H, CO}, i,j}$ and the $R_{\rm H, CO}$ of the corresponding single-binding model; each panel is a snapshot of one of the $(r, t)$ combinations indicated by the six circular markers in \fig{fig:tdust_map} and \fig{fig:flux_diff}. The upper-right panel shows the coldest case (15\,K), where the shape of the whitish area is determined by the interplay between the hopping terms (higher when \emph{at least} one of the two species has lower binding energies), the abundances of the species in each binding site (higher when \emph{both} species have higher binding energies), and the binding sites availability, i.e.~the distribution $P$ (higher when \emph{both} species have binding energies that corresponds to the centre of $P$).

In other words, this approach allows the existence of molecules bound to high-energy binding sites that react with hopping molecules, producing reactions that are not efficient in the classical single-binding scenario.

The combination of the three conditions described before not only determines the butterfly-shaped area at the centre of the upper-right panel (see \fig{fig:sketch_flux}), but also the similar features in the upper-middle and lower-right panels, where this effect is more prominent and shifted toward higher binding energies, because of the relatively higher dust temperature (35.4\,K and 38.0\,K). Note that, as the temperature increases, the absolute value of the reference $R_{\rm H, CO}$ decreases, as indicated by $R_{\rm ref}$ in each panel and by \fig{fig:flux_diff}, given the overall reduced amount of bound species.

When the temperature reaches 45.9\,K (upper-left panel), the features of the previous panels depend on the binding energy of CO only, since H has relatively lower binding energies than CO.

Finally, when the dust grains become hotter (lower-left and lower-middle panels, 132.1\,K and 102.9\,K, respectively), the effect is smoothed and also characterized by considerably smaller values of $R_{\rm ref}$, causing a less prominent divergence from the single-binding case, see also \fig{fig:flux_diff}.

Note that taking into account multiple bins might worsen the stochastic problem for reactions with H at $T>15-20$\,K, when the total concentration of hydrogen atoms per grain becomes $<1$, so that its concentration in a specific bin will be even smaller (e.g.~\citealt{Tielens1982} and \citealt{Caselli2002}). We therefore expect that in \fig{fig:flux_diff} the differences found for A and B, could be less prominent when using a more accurate treatment of stochasticity, via e.g.~a Monte Carlo method, however, with the approach employed in our paper (and commonly used in the astrophysical community), our conclusions remain unaffected by this specific problem, as the prefactor used to account for tunnelling of H does not depend on the binding energy. When thermal hopping is considered for H atoms, this problem is not relevant, as discussed in \citet{Katz1999} and \citet{Garrod2008}. Additionally, note that H\,+\,CO in this work is a proxy reaction, but the multiple binding energy approach is applicable to any X\,+\,Y surface products interaction, hence even to reactions without stochasticity issues.

\begin{table}
    \centering
	\begin{tabular}{lrlr}
		\hline
        Species & $n_i/n$ & Species & $n_i/n$\\
		\hline
        H & $10^{-3}$ & CO & $10^{-4}$\\
        H$_2$ & $5\times10^{-1}$ & H$_2$O & $3\times10^{-3}$\\
        He & $10^{-1}$ && \\
		\hline
	\end{tabular}
	\caption{Initial abundances from \citet{Rollig2007} employed in our model in units of gas number density $n$. Species not listed in this table are initially set to zero.}\label{tab:initial_abund}
\end{table}

\subsection{Case study 2: The midplane of a static circumstellar disk}
We apply our model to the midplane of a circumstellar disk, a denser environment when compared to the previous case, and where the temperature of the dust decreases with the distance from the star embedded in the system. We follow the Minimum Mass Solar Nebula model (MMSN, e.g.~\citealt{Min2011}), with the following density and temperature radial profiles, where $r$ denotes the distance from the central star,
\beqa
  \rho(r) &=& \mu\, m_{\rm p}\,n(r) = \frac{\Sigma_0}{H(r) \sqrt{2\pi}} \left(\frac{r}{\rm 1\,au}\right)^{-3/2}\\
  T_{\rm d}(r) &=& T(r) = T_0\, \left(\frac{r}{\rm 1\,au}\right)^{-1/2}\,,
\eeqa
with $\mu=2.34$, and a scale height
\beq
  H(r) = \frac{c_{\rm s}}{\Omega_{\rm K}}\,,
\eeq
where the speed of sound and the Keplerian angular frequency are respectively
\beq
  c_{\rm s} = \sqrt{\frac{k_{\rm B}T(r)}{\mu m_{\rm p}}}\qquad{\rm and}\qquad  \Omega_{\rm K} = \sqrt{\frac{G M_*}{r^3}}\,,
\eeq
assuming $M_*=1$\,M$_\odot$, $\Sigma_0=1700$\,g\,cm$^{-2}$, $T_0=200$\,K, and where $G$ is the gravitational constant. The dust is assumed to have the same properties as of \sect{sect:protostar}, while the initial abundances are the same as in  \tab{tab:initial_abund}, as well as the cosmic-ray ionization rate and the visual extinction set to $\zeta=5\times10^{-17}$\,s$^{-1}$ and $A_{\rm v}=30$\,mag, respectively.

\begin{figure}
 \includegraphics[width=\linewidth]{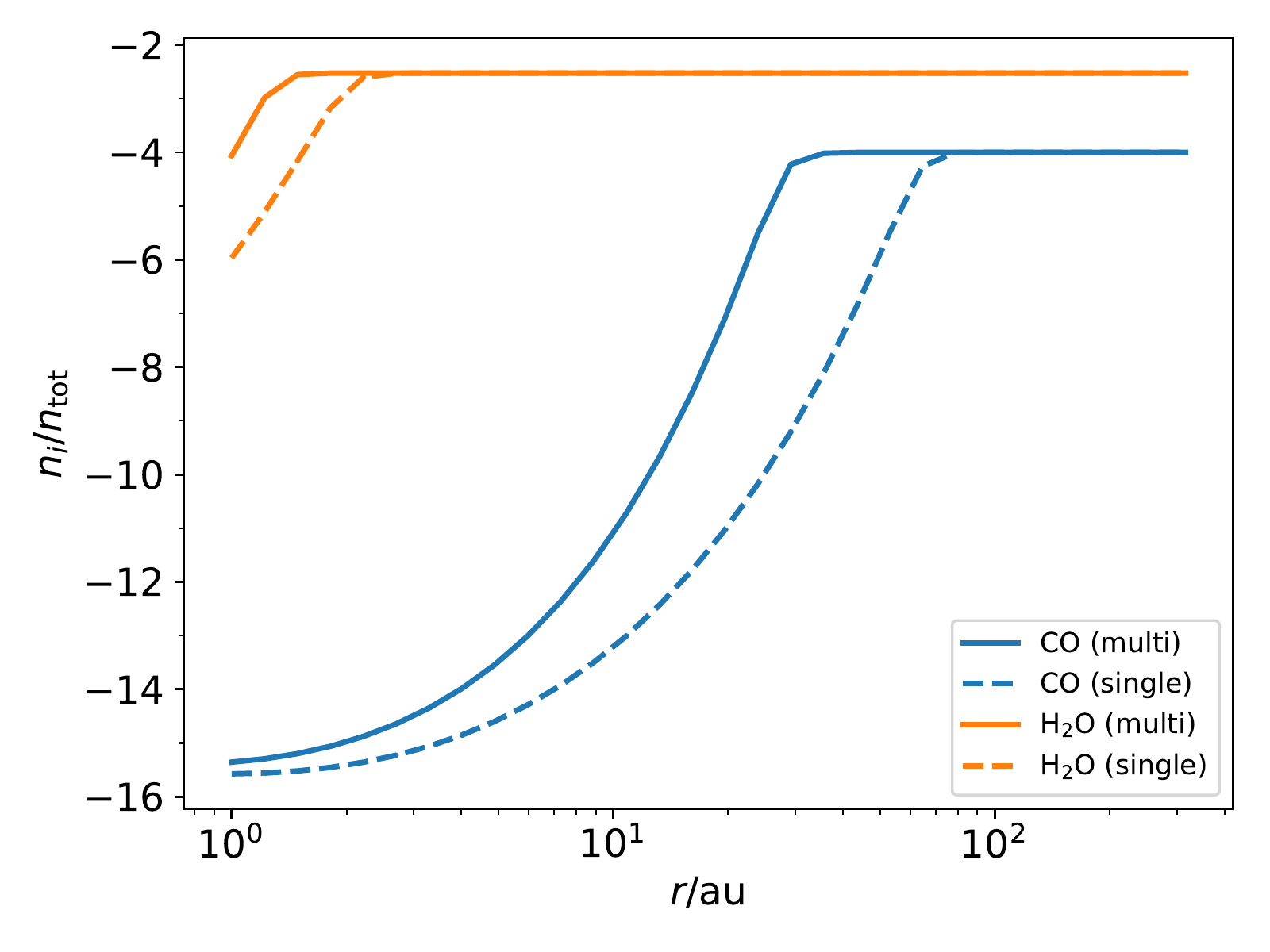}
 \caption{Comparison of the radial profiles of the abundances relative to the total density between single- (dashed) and multi-binding (solid) for CO (blue) and water (orange). The background model is the static MMSN protoplanetary disk described in the text.}\label{fig:disk_snowline}
\end{figure}

At each radius we let evolve the chemical abundances to equilibrium keeping the temperature and the density constant over time\footnote{The code to reproduce the disk model can be found at \url{https://bitbucket.org/tgrassi/multi\_bind/src/master/main_disk.py}}. In this scenario we are interested in determining the amount of CO and water condensed onto the dust grains at different positions of the disk. The outcome of the model is reported in \fig{fig:disk_snowline}, where the solid and the dashed lines indicate the abundance $n_{\dust{X}}=\sum_{i=1}^{N_{\rm b}} n_{\dust{X}, i}$ of CO (blue) and H$_2$O (orange) with the multi- and the single-binding approach, respectively.

In this scenario $T_{\rm d}$ decreases with $r$, hence the abundances of the species condensed onto the grain surface increase with $r$, and, analogously to the previous case, the multi-binding approach retains more molecules at relatively higher temperatures, given the availability of higher binding energy sites. This behaviour is shown in \fig{fig:disk_snowline}, where H$_2$O ice is formed around 1.3\,au in the multi-binding case and around 2\,au with single-binding, while CO is shifted from 90\,au to about 33\,au, aware that the exact values depend on $T_0$, that determines the temperature profile of the disk.

\subsection{Parameter analysis}\label{sect:parameters}
\begin{figure}
 \includegraphics[width=\linewidth]{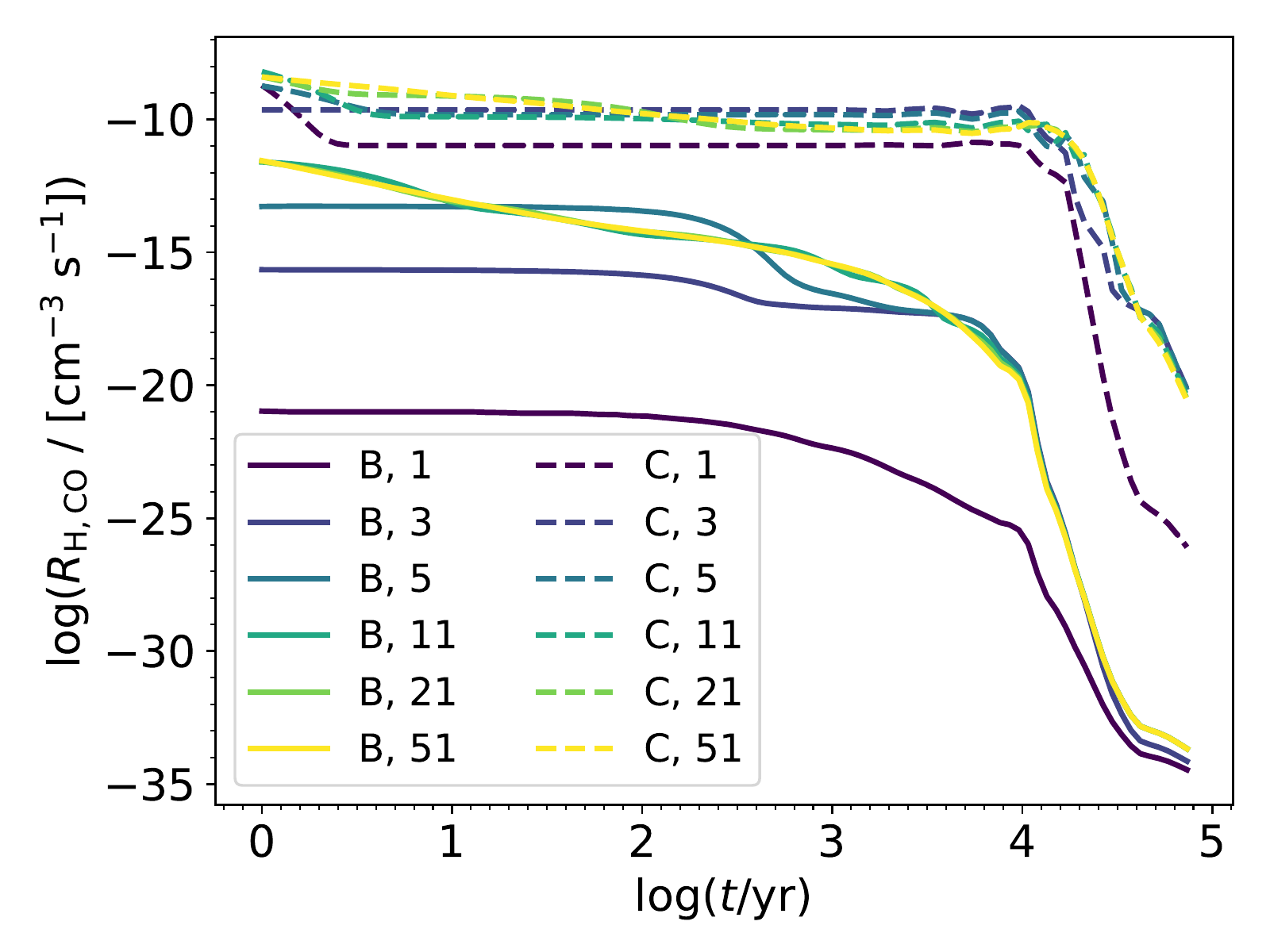}
 \caption{Cases B (solid) and C (dashed) as described in \fig{fig:flux_diff}. The different $N_{\rm b}$ in the legend are indicated with different colors. Case A is not reported for the sake of clarity, but presents similar convergence with $N_{\rm b}$ to case B.}\label{fig:nb_var}
\end{figure}

In order to avoid convergence problems, we performed our previous models with $N_{\rm b}=51$ bins of binding energy, however this number affects the overall efficiency of the chemical solver, since the number of each evaporation/adsorption reaction is increased by a factor $N_{\rm b}$, while the number of each surface reaction by a factor $N_{\rm b}^{1+M}$, where $M$ is the number of products. The effect of changing $N_{\rm b}$ is reported in \fig{fig:nb_var} by plotting the same quantity of \fig{fig:flux_diff} for the cases B and C, by changing the number of bins as $N_{\rm b}=[1, 5, 11, 21, 51]$. We note that $N_{\rm b}=21$ reproduces the results of $N_{\rm b}=51$, while $N_{\rm b}=11$ works for B, but not for C that presents some divergence. Case A (not reported here for the sake of clarity) shows the same behaviour as case B.

\begin{figure}
 \includegraphics[width=\linewidth]{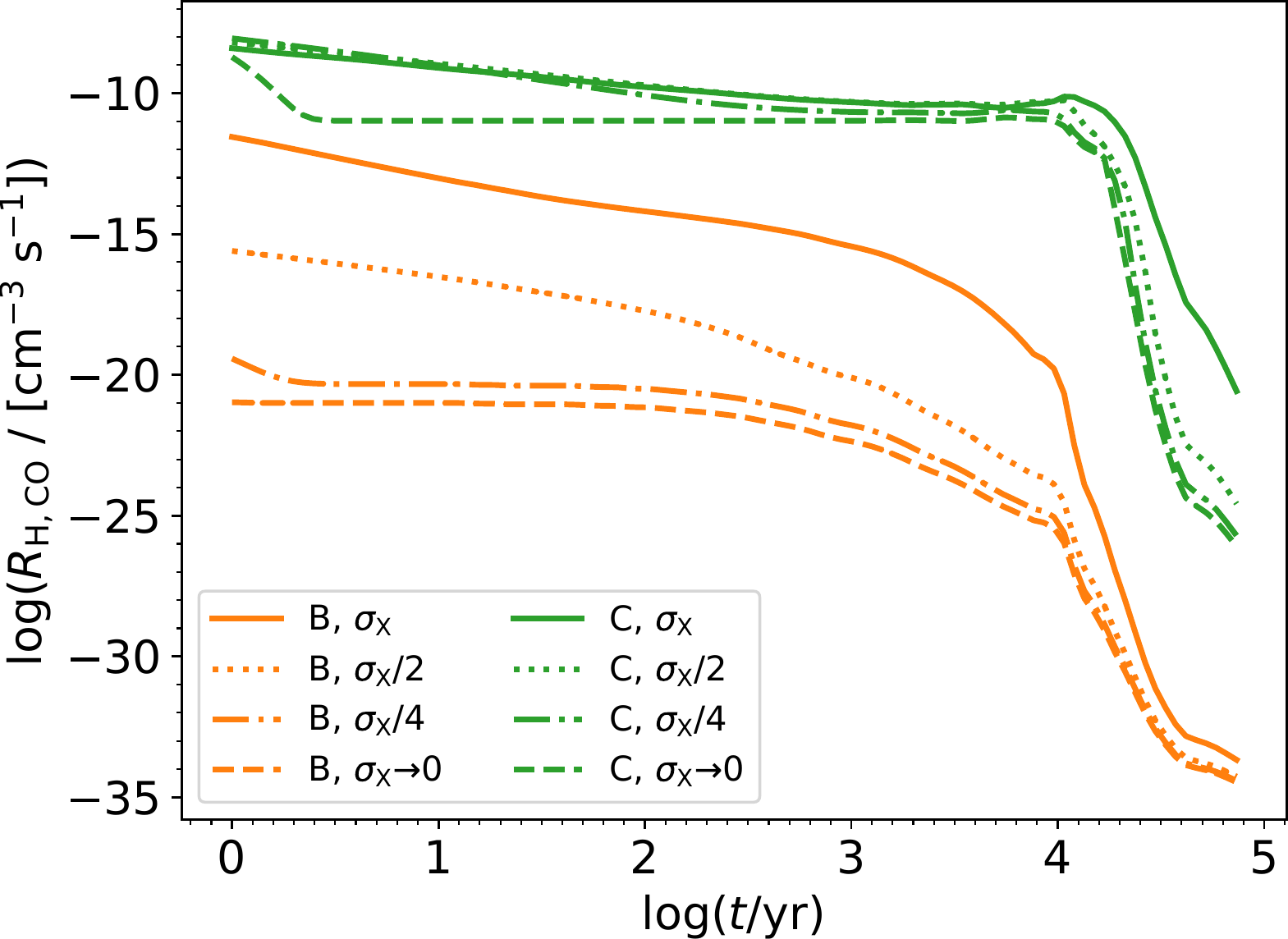}
 \caption{Cases B (orange) and C (green) as described in \fig{fig:flux_diff}. We compare $\sigma_{\rm X}$ as in \tab{tab:gauss} (solid), half (dotted), quarter (dash-dotted) of the value, and single case (dashed), i.e.~$\sigma_{\rm X}\to0$. Case A is not reported for the sake of clarity, but presents a similar behaviour to case B.}\label{fig:sigma_var}
\end{figure}

\fig{fig:sigma_var} shows the effect of changing $\sigma_{\rm X}$, in particular half ($\sigma_{\rm X}/2$, dotted) and a quarter ($\sigma_{\rm X}/4$, dash-dotted) of the original value (solid), as well as the single-binding ($\sigma_{\rm X}\to0$, dashed). As expected, reducing this value produces results that converge to the single-binding limiting case, i.e.~$\sigma_{\rm X}\to 0$, and $\sigma_{\rm X}/4$ shows a behaviour close to the single-binding case, suggesting that additional experiments and theoretical studies are necessary to determine $P(T_{\rm b})$ with accuracy.

\section{Discussion and Outlook}\label{sect:conclusions}
We have implemented a framework to explore the effects of a distribution of binding energies on the grain sites that participate to the chemistry on dust, rather than a single value, as generally employed in chemical models. This approach is supported by recent theoretical and experimental findings that show distributions resembling Gaussian functions.

Our results suggest that employing a distribution allows the molecules to have access to higher-energy binding sites, hence increasing their residence time onto the grain surface, and then becoming available for reacting with other molecules even at dust temperatures that usually present poor or no reactivity at all.

We also found that, given the dust temperature, the surface reactivity is affected by the interplay of three ingredients (cfr.~\fig{fig:sketch_flux}):
\begin{itemize}
 \item \emph{Residence time}: in the high-energy part of the binding sites distribution $P(T_{\rm b})$, molecules remains for longer on the grains, being available for reactions at higher temperatures.
 \item \emph{Thermal hopping}: on the other hand, the high-energy region of $P$ has lower hopping efficiency, hence reducing the reactivity.
 \item \emph{Sites availability}: even if there are combinations of reactants with long residence time and high hopping efficiency, the reactivity is ultimately determined by the wings of $P$, where there are (by construction) less available sites where to bound molecules.
\end{itemize}

Our models show that the combination of these three effects is relevant in astrophysical environments like the gas surrounding protostars and in protoplanetary disks, with consequences on the formation of interstellar complex organic molecules and on the location of the so-called snow-lines.

In the first case we followed the time-dependent evolution of a chemical network, computed alongside the variation of the dust temperature caused by the change in the protostar's luminosity. In particular, we followed the efficiency of the reaction \mbox{$\dust{H} + \dust{CO}\to products$}, finding differences that spans several orders of magnitude, depending on density and temperature.

Analogously, on the midplane of a protoplanetary disk with the multi-binding approach, molecules like CO and H$_2$O can be found on grains at a closer distance from the central star, where the dust is relatively warmer. This determines the position of the snow-lines, that play a key role in regulating the position and the characteristics of the planet-forming regions of the disk.

It is important to notice that our model follows the widely-used approach that does not make a distinction between the position of the monolayers in the ice mantle. However, if we assume that deeper layers behave differently, we might obtain different results when the ice thickness increases, given that the chemistry of the bulk ice depends on cracks, mantle porosity \citep{Mispelaer2013,Yoneda2016} and on the possible lack of bulk diffusion \citep{Ghesquiere2018,Shingledecker2019}.

In conclusion, by exploring a set of astrophysical models, we found that \emph{including a multi-binding framework into chemical models determines a substantial difference in their outcomes}. However, in a practical situation (i) the number of reactions needed for multi-binding is considerably large (affecting the computational efficiency of most of the state-of-the-art chemical models) and (ii) the exact shape of the binding energy distribution function will play a key role in the evolution of these chemical models.
These two points suggest that it is crucial to \emph{find affordable solutions in order to simplify the problem} from a numerical point of view and to \emph{increase the number of theoretical and experimental works} to constrain the uncertainties.
We also stress the need of systematic theoretical studies to build a proper database of accurate binding energies distributions, not only on ASW structure, but also on a mixture of ices and as a function of the coverage parameter.

\section*{Acknowledgements}
The authors want to thank the referees for their useful comments and insights.
SB is financially supported by CONICYT Fondecyt Iniciaci\'on (project 11170268), CONICYT programa de Astronomia Fondo Quimal 2017 QUIMAL170001, and BASAL Centro de Astrofisica y Tecnologias Afines (CATA) AFB-17002.
GB acknowledges support from Beca de Doctorado Nacional ANID n. 21200180.
SVG is financially supported by ANID Fondecyt Iniciacion (project 11170949).
BE acknowledges support from the DFG cluster of excellence ``Origin and Structure of the Universe''
(\verb+http://www.universe-cluster.de/+).
This work was funded by the DFG Research Unit FOR 2634/1 ER685/11-1.




\bibliographystyle{aa}
\bibliography{mybib} 



\begin{appendix}
\section{Chemical network}\label{sect:chemical_network}
The chemical reactions employed in our models are reported in \tab{table:network}, following \citet{Glover2010} as implemented in \citet{Grassi2017}. The rate coefficients in machine-readable format can be found at \url{https://bitbucket.org/tgrassi/multi\_bind/src/master/networks/}, and in \sect{sect:models} for surface reactions.

\begin{table*}
\begin{adjustwidth}{0cm}{}

\begin{tabular*}{1.4\textwidth}{llll}
\footnotesize
\centering
\setlength{\tabcolsep}{3pt}
\begin{tabular}{>{\bf}llll}
\hline
1 & H+H&$\to$&H$_2$\\
2 & H+e$^-$&$\to$&H$^+$+2e$^-$\\
3 & H$^+$+e$^-$&$\to$&H\\
4 & He+e$^-$&$\to$&He$^+$+2e$^-$\\
5 & He$^+$+e$^-$&$\to$&He\\
6 & He$^+$+H&$\to$&He+H$^+$\\
7 & He+H$^+$&$\to$&He$^+$+H\\
8 & H$_2$+He&$\to$&He+2H\\
9 & H$_2$+He$^+$&$\to$&He+H$_2$$^+$\\
10 & H$_2$+He$^+$&$\to$&He+H+H$^+$\\
11 & H$_2$+He$^+$&$\to$&He$^+$+2H\\
12 & H+e$^-$&$\to$&H$^-$\\
13 & H$^-$+H&$\to$&H$_2$+e$^-$\\
14 & H+H$^+$&$\to$&H$_2$$^+$\\
15 & H$_2$$^+$+H&$\to$&H$_2$+H$^+$\\
16 & H$_2$+H$^+$&$\to$&H$_2$$^+$+H\\
17 & H$_2$+e$^-$&$\to$&2H+e$^-$\\
18 & H$_2$+H&$\to$&3H\\
19 & H$^-$+e$^-$&$\to$&H+2e$^-$\\
20 & H$^-$+H&$\to$&2H+e$^-$\\
21 & H$^-$+H$^+$&$\to$&2H\\
22 & H$^-$+H$^+$&$\to$&H$_2$$^+$+e$^-$\\
23 & H$_2$$^+$+e$^-$&$\to$&2H\\
24 & H$_2$$^+$+H$^-$&$\to$&H+H$_2$\\
25 & H$_2$+H$_2$&$\to$&H$_2$+2H\\
26 & H+H+He&$\to$&H$_2$+He\\
27 & H+H+H&$\to$&H$_2$+H\\
28 & H$_2$+H+H&$\to$&2H$_2$\\
29 & C$^+$+e$^-$&$\to$&C\\
30 & O$^+$+e$^-$&$\to$&O\\
31 & C+e$^-$&$\to$&C$^+$+2e$^-$\\
32 & O+e$^-$&$\to$&O$^+$+2e$^-$\\
33 & O$^+$+H&$\to$&O+H$^+$\\
34 & O+H$^+$&$\to$&O$^+$+H\\
35 & O+He$^+$&$\to$&O$^+$+He\\
36 & C+H$^+$&$\to$&C$^+$+H\\
37 & C$^+$+H&$\to$&C+H$^+$\\
38 & C+He$^+$&$\to$&C$^+$+He\\
39 & OH+H&$\to$&O+2H\\
40 & HOC$^+$+H$_2$&$\to$&HCO$^+$+H$_2$\\
41 & HOC$^+$+CO&$\to$&HCO$^+$+CO\\
42 & C+H$_2$&$\to$&CH+H\\
43 & CH+H&$\to$&C+H$_2$\\
44 & CH+H$_2$&$\to$&CH$_2$+H\\
45 & CH+C&$\to$&C$_2$+H\\
46 & CH+O&$\to$&CO+H\\
47 & CH+O&$\to$&HCO$^+$+e$^-$\\
48 & CH+O&$\to$&OH+C\\
49 & CH$_2$+H&$\to$&CH+H$_2$\\
50 & CH$_2$+O&$\to$&CO+2H\\
51 & CH$_2$+O&$\to$&CO+H$_2$\\
52 & CH$_2$+O&$\to$&HCO+H\\
53 & CH$_2$+O&$\to$&CH+OH\\
54 & C$_2$+O&$\to$&CO+C\\
55 & O+H$_2$&$\to$&OH+H\\
56 & OH+H&$\to$&O+H$_2$\\
57 & H$_2$+OH&$\to$&H$_2$O+H\\
58 & C+OH&$\to$&H+CO\\
59 & O+OH&$\to$&H+O$_2$\\
60 & OH+OH&$\to$&H$_2$O+O\\
\hline
\end{tabular}

\begin{tabular}{>{\bf}llll}
\hline
61 & H$_2$O+H&$\to$&H$_2$+OH\\
62 & O$_2$+H&$\to$&OH+O\\
63 & O$_2$+H$_2$&$\to$&2OH\\
64 & O$_2$+C&$\to$&CO+O\\
65 & CO+H&$\to$&C+OH\\
66 & H$_2$$^+$+H$_2$&$\to$&H$_3$$^+$+H\\
67 & H$_3$$^+$+H&$\to$&H$_2$$^+$+H$_2$\\
68 & C+H$_2$$^+$&$\to$&CH$^+$+H\\
69 & C+H$_3$$^+$&$\to$&CH$^+$+H$_2$\\
70 & C+H$_3$$^+$&$\to$&CH$_2$$^+$+H\\
71 & C$^+$+H$_2$&$\to$&CH$^+$+H\\
72 & CH$^+$+H&$\to$&C$^+$+H$_2$\\
73 & CH$^+$+H$_2$&$\to$&CH$_2$$^+$+H\\
74 & CH$^+$+O&$\to$&CO$^+$+H\\
75 & CH$_2$$^+$+H&$\to$&CH$^+$+H$_2$\\
76 & CH$_2$$^+$+H$_2$&$\to$&CH$_3$$^+$+H\\
77 & CH$_2$$^+$+O&$\to$&HCO$^+$+H\\
78 & CH$_3$$^+$+H&$\to$&CH$_2$$^+$+H$_2$\\
79 & CH$_3$$^+$+O&$\to$&HOC$^+$+H$_2$\\
80 & CH$_3$$^+$+O&$\to$&HCO$^+$+H$_2$\\
81 & C$_2$+O$^+$&$\to$&CO$^+$+C\\
82 & O$^+$+H$_2$&$\to$&H+OH$^+$\\
83 & O+H$_2$$^+$&$\to$&H+OH$^+$\\
84 & O+H$_3$$^+$&$\to$&H$_2$+OH$^+$\\
85 & O+H$_3$$^+$&$\to$&H+H$_2$O$^+$\\
86 & OH+H$_3$$^+$&$\to$&H$_2$+H$_2$O$^+$\\
87 & OH+C$^+$&$\to$&H+CO$^+$\\
88 & OH$^+$+H$_2$&$\to$&H$_2$O$^+$+H\\
89 & H$_2$O$^+$+H$_2$&$\to$&H$_3$O$^+$+H\\
90 & H$_2$O+H$_3$$^+$&$\to$&H$_2$+H$_3$O$^+$\\
91 & H$_2$O+C$^+$&$\to$&HOC$^+$+H\\
92 & H$_2$O+C$^+$&$\to$&HCO$^+$+H\\
93 & H$_2$O+C$^+$&$\to$&H$_2$O$^+$+C\\
94 & H$_3$O$^+$+C&$\to$&HCO$^+$+H$_2$\\
95 & O$_2$+C$^+$&$\to$&CO$^+$+O\\
96 & O$_2$+C$^+$&$\to$&CO+O$^+$\\
97 & O$_2$+CH$_2$$^+$&$\to$&HCO$^+$+OH\\
98 & C+O$_2$$^+$&$\to$&O+CO$^+$\\
99 & C+O$_2$$^+$&$\to$&O$_2$+C$^+$\\
100 & CO+H$_3$$^+$&$\to$&H$_2$+HCO$^+$\\
101 & CO+H$_3$$^+$&$\to$&H$_2$+HOC$^+$\\
102 & HCO$^+$+C&$\to$&CO+CH$^+$\\
103 & HCO$^+$+H$_2$O&$\to$&CO+H$_3$O$^+$\\
104 & CH+H$^+$&$\to$&CH$^+$+H\\
105 & CH$_2$+H$^+$&$\to$&H$_2$+CH$^+$\\
106 & CH$_2$+H$^+$&$\to$&H+CH$_2$$^+$\\
107 & CH$_2$+He$^+$&$\to$&He+H$_2$+C$^+$\\
108 & CH$_2$+He$^+$&$\to$&He+H+CH$^+$\\
109 & C$_2$+He$^+$&$\to$&C$^+$+C+He\\
110 & OH+H$^+$&$\to$&OH$^+$+H\\
111 & OH+He$^+$&$\to$&O$^+$+He+H\\
112 & H$_2$O+H$^+$&$\to$&H+H$_2$O$^+$\\
113 & H$_2$O+He$^+$&$\to$&He+OH+H$^+$\\
114 & H$_2$O+He$^+$&$\to$&He+OH$^+$+H\\
115 & H$_2$O+He$^+$&$\to$&He+H$_2$O$^+$\\
116 & O$_2$+H$^+$&$\to$&O$_2$$^+$+H\\
117 & O$_2$+He$^+$&$\to$&O$_2$$^+$+He\\
118 & O$_2$+He$^+$&$\to$&O$^+$+He+O\\
119 & CO+He$^+$&$\to$&C$^+$+He+O\\
120 & CO+He$^+$&$\to$&C+He+O$^+$\\
\hline
\end{tabular}

\begin{tabular}{>{\bf}llll}
\hline
121 & CO$^+$+H&$\to$&CO+H$^+$\\
122 & C$^-$+H$^+$&$\to$&C+H\\
123 & O$^-$+H$^+$&$\to$&O+H\\
124 & He$^+$+H$^-$&$\to$&H+He\\
125 & H$_3$$^+$+e$^-$&$\to$&H$_2$+H\\
126 & H$_3$$^+$+e$^-$&$\to$&3H\\
127 & CH$^+$+e$^-$&$\to$&C+H\\
128 & CH$_2$$^+$+e$^-$&$\to$&CH+H\\
129 & CH$_2$$^+$+e$^-$&$\to$&C+H$_2$\\
130 & CH$_2$$^+$+e$^-$&$\to$&C+2H\\
131 & CH$_3$$^+$+e$^-$&$\to$&CH$_2$+H\\
132 & CH$_3$$^+$+e$^-$&$\to$&CH+H$_2$\\
133 & CH$_3$$^+$+e$^-$&$\to$&CH+2H\\
134 & OH$^+$+e$^-$&$\to$&O+H\\
135 & H$_2$O$^+$+e$^-$&$\to$&O+H$_2$\\
136 & H$_2$O$^+$+e$^-$&$\to$&OH+H\\
137 & H$_2$O$^+$+e$^-$&$\to$&O+2H\\
138 & H$_3$O$^+$+e$^-$&$\to$&OH+2H\\
139 & H$_3$O$^+$+e$^-$&$\to$&O+H+H$_2$\\
140 & H$_3$O$^+$+e$^-$&$\to$&H+H$_2$O\\
141 & H$_3$O$^+$+e$^-$&$\to$&OH+H$_2$\\
142 & O$_2$$^+$+e$^-$&$\to$&2O\\
143 & CO$^+$+e$^-$&$\to$&C+O\\
144 & HCO$^+$+e$^-$&$\to$&CO+H\\
145 & HCO$^+$+e$^-$&$\to$&OH+C\\
146 & HOC$^+$+e$^-$&$\to$&CO+H\\
147 & H$^-$+C&$\to$&CH+e$^-$\\
148 & H$^-$+O&$\to$&OH+e$^-$\\
149 & H$^-$+OH&$\to$&H$_2$O+e$^-$\\
150 & C$^-$+H&$\to$&CH+e$^-$\\
151 & C$^-$+H$_2$&$\to$&CH$_2$+e$^-$\\
152 & C$^-$+O&$\to$&CO+e$^-$\\
153 & O$^-$+H&$\to$&OH+e$^-$\\
154 & O$^-$+H$_2$&$\to$&H$_2$O+e$^-$\\
155 & O$^-$+C&$\to$&CO+e$^-$\\
156 & H$_2$+H$^+$&$\to$&2H+H$^+$\\
157 & H$_2$+H$^+$&$\to$&H$_3$$^+$\\
158 & C+e$^-$&$\to$&C$^-$\\
159 & C+H&$\to$&CH\\
160 & C+H$_2$&$\to$&CH$_2$\\
161 & C+C&$\to$&C$_2$\\
162 & C+O&$\to$&CO\\
163 & C$^+$+H&$\to$&CH$^+$\\
164 & C$^+$+H$_2$&$\to$&CH$_2$$^+$\\
165 & C$^+$+O&$\to$&CO$^+$\\
166 & O+e$^-$&$\to$&O$^-$\\
167 & O+H&$\to$&OH\\
168 & O+O&$\to$&O$_2$\\
169 & OH+H&$\to$&H$_2$O\\
\hline
170 & H$^-$+$\gamma$&$\to$&H+e$^-$\\
171 & H$_2$$^+$+$\gamma$&$\to$&H+H$^+$\\
172 & H$_3$$^+$+$\gamma$&$\to$&H$_2$+H$^+$\\
173 & H$_3$$^+$+$\gamma$&$\to$&H$_2$$^+$+H\\
174 & C+$\gamma$&$\to$&C$^+$+e$^-$\\
175 & C$^-$+$\gamma$&$\to$&C+e$^-$\\
176 & CH+$\gamma$&$\to$&C+H\\
177 & CH+$\gamma$&$\to$&CH$^+$+e$^-$\\
178 & CH$^+$+$\gamma$&$\to$&C+H$^+$\\
179 & CH$_2$+$\gamma$&$\to$&CH+H\\
180 & CH$_2$+$\gamma$&$\to$&CH$_2$$^+$+e$^-$\\
\hline
\end{tabular}

\begin{tabular}{>{\bf}llll}
\hline
181 & CH$_2$$^+$+$\gamma$&$\to$&CH$^+$+H\\
182 & CH$_3$$^+$+$\gamma$&$\to$&CH$_2$$^+$+H\\
183 & CH$_3$$^+$+$\gamma$&$\to$&CH$^+$+H$_2$\\
184 & C$_2$+$\gamma$&$\to$&2C\\
185 & O$^-$+$\gamma$&$\to$&O+e$^-$\\
186 & OH+$\gamma$&$\to$&O+H\\
187 & OH+$\gamma$&$\to$&OH$^+$+e$^-$\\
188 & OH$^+$+$\gamma$&$\to$&O+H$^+$\\
189 & H$_2$O+$\gamma$&$\to$&OH+H\\
190 & H$_2$O+$\gamma$&$\to$&H$_2$O$^+$+e$^-$\\
191 & O$_2$+$\gamma$&$\to$&O$_2$$^+$+e$^-$\\
192 & O$_2$+$\gamma$&$\to$&2O\\
193 & CO+$\gamma$&$\to$&C+O\\
194 & H$_2$+$\gamma$&$\to$&2H\\
195 & H$_2$O$^+$&$\to$&H$_2$$^+$+O\\
196 & H$_2$O$^+$&$\to$&H$^+$+OH\\
197 & H$_2$O$^+$&$\to$&O$^+$+H$_2$\\
198 & H$_2$O$^+$&$\to$&OH$^+$+H\\
199 & H$_3$O$^+$&$\to$&H$^+$+H$_2$O\\
200 & H$_3$O$^+$&$\to$&H$_2$$^+$+OH\\
201 & H$_3$O$^+$&$\to$&H$_2$O$^+$+H\\
202 & H$_3$O$^+$&$\to$&OH$^+$+H$_2$\\
\hline
203 & H+CR&$\to$&H$^+$+e$^-$\\
204 & He+CR&$\to$&He$^+$+e$^-$\\
205 & O+CR&$\to$&O$^+$+e$^-$\\
206 & CO+CR&$\to$&C+O\\
207 & CO+CR&$\to$&CO$^+$+e$^-$\\
208 & C$_2$+CR&$\to$&2C\\
209 & H$_2$+CR&$\to$&2H\\
210 & H$_2$+CR&$\to$&H$^+$+H$^-$\\
211 & H$_2$+CR&$\to$&H$_2$$^+$+e$^-$\\
212 & C+CR&$\to$&C$^+$+e$^-$\\
213 & CH+CR&$\to$&C+H\\
214 & O$_2$+CR&$\to$&2O\\
215 & O$_2$+CR&$\to$&O$_2$$^+$+e$^-$\\
216 & OH+CR&$\to$&O+H\\
217 & CH$_2$+CR&$\to$&CH$_2$$^+$+e$^-$\\
218 & H$_2$O+CR&$\to$&OH+H\\
219 & HCO+CR&$\to$&CO+H\\
220 & HCO+CR&$\to$&HCO$^+$+e$^-$\\
221 & H$_2$+CR&$\to$&H+H$^+$+e$^-$\\
\hline
222 & C+C + H$_2$ &$\to$&C$_2$ + H$_2$\\
223 & C+O + H$_2$ &$\to$&CO + H$_2$\\
224 & C+O$^+$ + H$_2$ &$\to$&CO$^+$ + H$_2$\\
225 & C$^+$+O + H$_2$ &$\to$&CO$^+$ + H$_2$\\
226 & H+O + H$_2$&$\to$&OH + H$_2$\\
227 & OH+O + H$_2$&$\to$&H$_2$O + H$_2$\\
228 & O+O + H$_2$&$\to$&O$_2$ + H$_2$\\
\hline
229 & H & $\to$ & $\dust{H}$\\
230 & CO & $\to$ & $\dust{CO}$\\
231 & H$_2$O & $\to$ & $\dust{H_2O}$\\
232 & $\dust{H}$ & $\to$ & H\\
233 & $\dust{CO}$ & $\to$ & CO\\
234 & $\dust{H_2O}$ & $\to$ & H$_2$O\\
235 & $\dust{H}$ + $\dust{CO}$ & $\to$ & $products$\\
\hline
\phantom{999}&&&\\
\phantom{999}&&&\\
\phantom{999}&&&\\
\phantom{999}&&&\\
\phantom{999}&&&\\
\end{tabular}
\end{tabular*}
\end{adjustwidth}
\caption{Chemical network employed in this paper. Photochemical reactions includes photons $\gamma$, while cosmic rays are indicated with CR. Note that the first reaction represents H$_2$ catalysis on grains, e.g.~\citet{Hollenbach1979}. More details can be found in \citet{Glover2010} and \citet{Grassi2017} and in the text. See \sect{sect:models} for additional information about surface reactions.}\label{table:network}
\end{table*}

\end{appendix}

\end{document}